\documentstyle[12pt]{article}

\makeatletter
\@addtoreset{equation}{section}
\makeatother

\newcommand{\be}{\begin{equation}}
\newcommand{\ee}{\end{equation}}
\newcommand{\bee}{\begin{eqnarray}}
\newcommand{\eee}{\end{eqnarray}}

\newcommand{\ga}{\alpha}
\newcommand{\gb}{\beta}
\newcommand{\gga}{\gamma}
\newcommand{\gd}{\delta}
\newcommand{\gl}{\lambda}

\newcommand{\gep}{\epsilon}

\newcommand{\gs}{\sigma}
\newcommand{\go}{\omega}
\newcommand{\gt}{\tau}

\newcommand{\nn}{\nonumber}

\newcommand{\ptl}{\partial}
\newcommand{\ep}{\mbox{$\wedge$}}

\newcommand{\un}{{\underline{n}}}
\newcommand{\um}{{\underline{m}}}

\newcommand{\uk}{{\underline{k}}}

\renewcommand{\L}{{\cal L}}
\renewcommand{\P}{{\cal P}}
\newcommand{\K}{{\cal K}}
\newcommand{\D}{{\cal D}}
\newcommand{\R}{{\cal R}}
\newcommand{\Q}{{\cal Q}}
\renewcommand{\S}{{\cal S}}

\newcommand{\J}{{\cal J}}

\newcommand{\ket}{{\rangle}}

\newcommand{\rv}{{|0\ket}}
\newcommand{\novx}{{(a^+,a,k|x)}}
\newcommand{\nov}{{(a^+,a,k)}}
\newcommand{\vac}
{|0\rangle\langle 0|}
\newcommand{\uvac}
{|0_u\rangle\langle 0_u|}
\newcommand{\fv}
{|\Phi(a^+|x)\rangle}

\begin{document}

\begin{flushright}
\vspace{1mm}
FIAN/TD/07--01\\
{March 2001}\\
\end{flushright}

\vspace{1cm}

\begin{center}
{\large\bf Higher Spin Conformal Symmetry  for Matter Fields
in $2+1$ Dimensions}\\
\vglue 1  true cm
\vspace{1cm}
{\bf O.V.~Shaynkman
\footnote{e-mail: shayn@lpi.ru}
  and M.A.~Vasiliev }\footnote{e-mail: vasiliev@lpi.ru}  \\
\vspace{1cm}

I.E.Tamm Department of Theoretical Physics, Lebedev Physical
Institute,\\
Leninsky prospect 53, 117924, Moscow, Russia
\vspace{1.5cm}
\end{center}

\begin{abstract}
A simple realization of the conformal
higher spin symmetry on the free
$3d$ massless matter fields is given
in terms of an auxiliary Fock module
both in the flat and $AdS_3$ case. The duality between
non-unitary field-theoretical representations of the
conformal algebra and
the unitary (singleton--type) representations of the $3d$ conformal
algebra $sp(4,\R)$ is formulated explicitly in terms of a certain
Bogolyubov transform.
\end{abstract}

\section{Introduction}
The AdS/CFT correspondence \cite{Dirac,AdS/CFT} suggests duality
between theories of gravity in the bulk $AdS_d$ space and
$d-1$ dimensional conformal theories at the boundary of $AdS_d$.
Some time ago a consistent theory of interacting massless fields
of all spins in $AdS_4$ was developed
\cite{more} (see also \cite{Gol,qg} for reviews and more
references). As the higher spin gauge theory contains gravity
and supergravity one may speculate that it should
have some conformal dual
theory exhibiting the infinite-dimensional $AdS_4$ higher
spin symmetries
as $3d$ conformal higher spin symmetries. In fact, the $3d$
conformal higher spin symmetry was identified
long ago  \cite{FL} with the $AdS_4$ higher spin algebra
\cite{FV,Fort2}, thus suggesting the higher spin
AdS/CFT correspondence. In the same reference \cite{FL}
the set of $3d$ conformal higher spin gauge fields associated with
the conformal higher spin algebra was introduced. However, $3d$
higher spin gauge fields do not propagate, i.e. their own
dynamics is topological\footnote{The terminology ``higher spin"
is therefore somewhat sloppy. It means  that the higher spin
gauge fields and symmetry parameters are tensors of higher ranks.}.
To find a nontrivial boundary conformal theory one
has to realize the $3d$ conformal higher spin symmetries on
some matter fields. This is the goal of the present paper.
We show that the conformal higher spin symmetry admits a natural
realization on the $3d$ massless scalar and spinor. This realization
is given in terms of an auxiliary Fock space dual to
the singleton representation in the unitary Fock space \cite{Dirac}
by a certain Bogolyubov transform.
Let us note that the suggested realization is different {}from
that used to describe $3d$ higher spin theories in \cite{V,PV}.

\section{3d Conformal Higher Spin Algebra}
\label{Higher Spin Algebras}

Extending the usual correspondence between conformal symmetries
in $d$ dimensions and AdS symmetries in $d+1$
dimensions, $3d$ conformal higher spin superalgebras were
identified in \cite{FL} with the $4d$ AdS higher spin algebras
\cite{FV,Fort2} according to the following construction.
Consider the oscillators $\hat a_\ga$ and $\hat a^{+\gb}$ with the
commutation relations
\be
[\hat a_\ga,\hat a^{+\gb} ]=\gd_\ga^\gb\,,\qquad
[\hat a_\ga,\hat a_\gb ]=[\hat a^{+\ga},\hat a^{+\gb}]=0\,,
\label{oscil}
\ee
where $\ga,\gb=1,2$.
The isomorphism $so(3,2)\sim sp(4,\R)$ allows one
to use symplectic realization for $3d$ conformal (and $AdS_4$)
symmetries. The algebra $sp(4,\R)$ admits the standard
oscillator realization in terms of bilinears of the oscillators
(\ref{oscil}).
The isomorphism $so(3,2)\sim sp(4,\R)$ is expressed by the relations
$$
\L_{nm}=\frac{1}{2}\gep_{nmk}\gs^{k\ga}{}_{\gb} \L_{\ga}{}^{\gb}\,,\qquad
\L_{\ga}{}^{\gb} = \frac{1}{2}\Big (\hat a_\ga \hat a^{+\gb}
+\hat a^{+\gb} \hat a_\ga \Big ) -\frac{1}{4} \delta_\ga^\gb
\Big (\hat a_\gga \hat a^{+\gga}+\hat a^{+\gga}\hat a_{\gga}\Big )\,,
$$
\be
\P_n=\gs_n{}^{\ga\gb} \P_{\ga\gb}\,,\qquad \P_{\ga\gb}=
\frac{1}{2}\hat a_\ga \hat a_\gb\,,
\label{confg}
\ee
$$
\K_n=\gs_n{}_{\ga\gb} \K^{\ga\gb}\,,\qquad \K^{\ga\gb}=
\frac{1}{2}\hat a^{+\ga} \hat a^{+\gb}\,,
$$
$$
\D=\frac{1}{4}(\hat a_\ga \hat a^{+\ga}+\hat a^{+\ga}\hat a_{\ga})\,.
$$
Here $n,m=0,1,2$ are fiber indices which
are raised and lowered by the mostly minus
Minkowski metric $\eta_{nm}$ and
$\gs_n{}^{\ga\gb}=\gs_n{}^{\gb\ga}=(I,\;\gt_1,\;\gt_3)^{\ga\gb}$ where
$\gt_1^{\ga\gb},\;\gt_3^{\ga\gb}$ are the Pauli matrices.
Spinorial indices
$\ga,\gb$ are
raised and lowered by
$\gep_{\ga\gb}=-\gep_{\gb\ga}\, (\gep_{12}=\gep^{12}=1)$ as
$c^\ga=\gep^{\ga\gb}c_\gb,\;\; c_\gb=\gep_{\ga\gb}c^\ga$
($\gep_{\ga\gb}$ is the symplectic form of
$sp(2,\R)\sim so(2,1)$).
$\gs_n{}^{\ga\gb}$ satisfy the following identities
\be
\gs_n{}^{\ga\gb}\gs_m{}_{\ga\gb}=2\eta_{nm}\,,
\qquad \gs_n{}^{\ga\gb}\gs^n{}_{\ga'\gb'}=\gd_{\ga'}^\ga\gd_{\gb'}^\gb
+\ga\leftrightarrow\gb\,,
\ee
$$
\gs_n{}^{\ga}{}_{\gb}\gs_m{}^{\gb\gga}=\eta_{nm}\gep^{\ga\gga}-
\gep_{nmk}\gs^{k\ga\gga}\,.
$$

$\L_{\ga}{}^{\gb}$, $\P_{\ga\gb}$, $\K^{\ga\gb}$ and $\D$
identify, respectively, with the generators of the Lorentz rotations,
Poincare translations, special conformal transformations and
dilatations. The $N=2$ supersymmetric extension $osp(2|4,\R)$ of
$sp(4,\R)$ admits the following useful realization
\cite{FrVa}. One extends
(\ref{oscil}) by adding a new  generating element
$\hat k$ satisfying the relations
\bee
\hat k^2=\hat 1\,,\qquad
\{\hat k,\hat a_\ga\}=\{\hat k,\hat a^{+\ga}\}=0\,.
\label{klein}
\eee
The supergenerators $\Q_{j\ga}$, $\S_j{}^\ga$
and the $u(1)$ charge $\J$ are
\be
\Q_{j\ga}=\frac{1}{\sqrt{2}}
(i\hat k)^j \hat a_\ga\,,\qquad
\S_j{}^\ga=\frac{1}{\sqrt{2}}
(i\hat k)^j \hat a^{+\ga}\,,\qquad j=0 \quad
\mbox{or} \quad 1\,,
\label{oddg}
\ee
\be
\J=\frac{1}{4i}\hat k\,.
\label{eg}
\ee

Let the (associative)
enveloping algebra of the relations (\ref{oscil}),
(\ref{klein}) be denoted $AK_2$. The higher spin Lie
superalgebra
$hgl(1;1|4)$ is defined as the same linear space endowed
with the product law defined via the (anti)commutators
in $AK_2$, $[\hat f,\hat g]_\pm=\hat f\hat g -
(-1)^{\pi(\hat f)\pi(\hat g)} \hat g\hat f$. The canonical
$Z_2$ grading chosen in accordance with the standard relation
between spin and statistics counts oddness of a number of
spinor indices
\be
\hat f(-\hat a^+,-\hat a,\hat k) = (-1)^{\pi (\hat f)}
\hat f(\hat a^+,\hat a,\hat k)\,.
\ee

Left Fock module of the algebras $hgl(1;1|4)$  and $AK_2$ can be defined
by the relations
\be
\hat a_\ga \rv=0\,,\qquad \hat k\rv=\rv\,.
\label{akv}
\ee
The basis vectors are
\be
\hat a^{+\ga_1}\cdots \hat a^{+\ga_l}\rv\,.
\label{apv}
\ee

In practice, instead of working with the operator realization of
the algebra
$AK_2$, it is convenient to use its star product version
\be
\label{prod}
\!\!(f*g)(a^+\!,a)\!=\!\!\frac{1}{\pi^4}\!
\int\!\!
d^2 u d^2 v d^2 u^+ d^2 v^+\!
\exp(2v_\ga u^{+\ga}\!-2u_\ga v^{+\ga})
 f(a^+\!+u^+\!,a+u)g(a^+\!+v^+\!,a+v)\,,
\label{ints}
\ee
with
$f(a^+,a)$ and $g(a^+,a)$ being functions (polynomials or formal power
series)
of the commuting variables
$a^{+\ga}$ and  $a_\gb$.\footnote{The integral
is defined  so that
$
\frac{1}{\pi^2}\int d^2v d^2u^+\exp(2v_\ga u^{+\ga})=\frac{1}{\pi^2}
\int d^2u d^2v^+\exp(-2u_\ga v^{+\ga})=1
$
and
\bee
&\int d^2v d^2u^+\frac{\ptl}{\ptl v_\ga}
\Big( \exp(2v_\ga u^{+\ga}) f(v,u^+)\Big)=
\int d^2v d^2u^+\frac{\ptl}{\ptl u^+_\ga}
\Big( \exp(2v_\ga u^{+\ga}) f(v,u^+)\Big)=\nn\\
&\int d^2u d^2v^+\frac{\ptl}{\ptl u_\ga}
\Big( \exp(-2u_\ga v^{+\ga}) g(u,v^+)\Big)=
\int d^2u d^2v^+\frac{\ptl}{\ptl v^+_\ga}
\Big( \exp(-2u_\ga v^{+\ga}) g(u,v^+)\Big)=0\nn
\eee
for any polynomials $f(v,u^+)$ and $g(u,v^+)$. If someone prefers
a definition in terms of oscillating exponentials, this can easily be
achieved by an appropriate redefinition of the variables.
}

This formula describes the associative algebra with the defining
relations
\bee
&{[}a_\ga,a^{+\gb}{]}_*=\gd_\ga^\gb\,,\nn\\
&{[}a_\ga,a_\gb{]}_*={[}a^{+\ga},a^{+\gb}{]}_*=0\,.
\eee
The star product defined this way describes the product of symmetrized
polynomials of oscillators in terms of symbols of
operators and realizes the subalgebra $A_2 \subset AK_2$ spanned by the
$\hat k$ - independent elements.
The full algebra $AK_2$ is defined for the
functions of the form $f (a^+,a,k) = f_0 (a^+,a)+f_1 (a^+,a)k$
by the formula (\ref{ints})
along with
\bee
k*f (a^+,a,k) = f (-a^+,-a,k)*k=f(-a^+,-a,k)k \,,\qquad k*k=1 \,.
\eee
The algebra $AK_2$ admits the involution\footnote{Recall
that involution  is
an involutive semilinear antiautomorphism, i.e.
$\mu(f*g) = \mu (g)*\mu (f)$, $\mu^2=1$,
$\mu (\ga f )= \bar{\ga} \mu (f)$ ($\bar{\ga } $ is complex conjugated
to $\ga$). A conjugation is
an involutive semilinear automorphism, i.e.
$\gs(f*g) =\gs(f)*\gs(g)$, $\gs^2=1$,
$\gs (\ga f )= \bar{\ga} \gs (f)$. Any conjugation $\gs$
singles out a real form of a complex algebra by  the condition
$\gs (f) = f$.}
$\mu$
defined by the relations
\be
\label{inv}
\mu (a_\ga)=i a_\ga\,,\qquad \mu (a^{+\ga})=i a^{+\ga}
\,,\qquad \mu (k) = k\,.
\ee
Any involution $\mu$ of an associative algebra $A$ induces a
conjugation
\be
\gs (f) = -(i)^{\pi (f)} \mu (f)
\label{conj}
\ee
of the Lie superalgebra $L_A$ built {}from $A$
by virtue of (anti)commutators.
The algebra $hu(1;1|4)$ is the real form of the algebra $hgl(1;1|4)$
singled out by the
conjugation (\ref{conj}), i.e. a generic element
of $hu(1;1|4)$
$f^\R(a^+,a,k)=f_0^\R(a^+,a)+f_1^\R(a^+,a)k$
satisfies
\be
f_0^\R(a^+,a)=-(i)^{\pi(f^\R_0)}\overline{f_0^\R(-ia^+,-ia)}\,,\qquad
f_1^\R(a^+,a)=-(i)^{\pi(f^\R_1)}\overline{f_1^\R(ia^+,ia)}\,,
\ee
where bar denotes the complex conjugation of the power series
expansion coefficients.
Note that the superconformal generators (\ref{confg}), (\ref{oddg}),
(\ref{eg})
satisfy the reality conditions $\gs(f)=f$ and thus belong to
$hu(1;1|4)$.

Fock representation of $AK_2$ is its left module  $F^l$ spanned
by the vectors of the form
\be
f * \vac \,,
\label{fr}
\ee
where  $\vac$ is the projector to the vacuum space.
It is well-known (see e.g.
\cite{d2,Gopuk}) that Fock projectors admit
the exponential realization in the star product algebra.
The Fock vacuum satisfying (\ref{akv}) is
\be
\vac =2(1+k)\exp(-2a_\ga a^{+\ga}) \,.
\ee
Indeed, it is easy to see that
\be
a_\ga * \vac =\vac *a^{+\ga} =
0\,,\quad k*\vac =\vac * k= \vac\,,\quad \vac *\vac = \vac\,.
\ee
The right Fock module $F^r$
of $AK_2$ is spanned by the vectors $\vac * f\,.$
The important properties of the Fock vacuum $\vac$ are that it is
Lorentz invariant
\be
\label{linv}
\L_\ga{}^\gb * \vac = \vac * \L_\ga{}^\gb =0
\ee
and has the definite conformal weight 1/2
\be
\label{conv}
\D* \vac =\vac *\D = \frac{1}{2}\vac\,.
\ee

Note, that the $\mu$-conjugated Fock module $\mu (\vac ) =
2(1+k)\exp(2 a_\ga a^{+\ga})$
is different {}from $\vac$. Moreover, it belongs to a distinct
sector of the star product algebra because
$\vac * \mu (\vac) = \infty$. This fact is not important
{}from the perspective of the present paper in which only free
fields are considered  (i.e., matter field modules are not
multiplied), but it should be taken into account
when thinking of a nonlinear theory describing interactions
that would exhibit higher spin conformal symmetries.

\section{Conformally Invariant Vacua}
\label{V}

Let $\go(x)$ be a 1-form  taking values
in the higher spin algebra $AK_2$, i.e. $\go$ is the generating function
of the conformal higher spin gauge fields
\be
\go(x)=\sum_{q=0,1}\sum_{l,r=0}^\infty
\frac{1}{l!r!}\go_q(x)_{\ga_1\ldots \ga_l,}{}^{\gb_1 \ldots \gb_r}
a^{+\ga_1}\cdots
a^{+\ga_l} a_{\gb_1}\cdots a_{\gb_r}(k)^q\,.
\ee

The zero-curvature equation
\be
d\go=\go\ep *\go\,
\label{ezc}
\ee
is invariant under the higher spin conformal
gauge transformations
\be
\gd\go=d\gep-{[}\go,\gep{]}_*\,,
\label{gtw}
\ee
where $\gep\novx$ is an infinitesimal gauge symmetry parameter
and $d=dx^\un \frac{\partial}{\partial x^\un}$ (underlined indices
$\um$, $\un =0,1,2$ are used for
the components of differential forms).
Any vacuum solution $\go_0$ of the equation (\ref{ezc}) breaks the
local higher spin symmetry to its stability subalgebra with the
infinitesimal parameters $\gep_0\novx$ satisfying the equation
\be
d\gep_0-{[}\go_0 ,\gep_0{]}_*=0\,.
\label{ipgs}
\ee
The consistency of this equation is guaranteed by (\ref{ezc}).

Locally, the equation (\ref{ezc})
admits a pure gauge solution
\be
\go_0 =-g^{-1}* dg\,.
\label{wvg}
\ee
Here $g\novx$ is some invertible element of the algebra
$AK_2$, i.e.
$g^{-1}*g=g*g^{-1}=1$.
For $\go_0$  (\ref{wvg}) one finds that the
generic solution of (\ref{ipgs}) is
\be
\gep_0 (x)=g^{-1}(x) *\xi*g(x)\,,
\label{e0}
\ee
where $\xi\nov=\xi_0(a^+,a)+\xi_1(a^+,a)k$
is an arbitrary $x$--independent element playing a role of
the initial data for the equation (\ref{ipgs})
\be
\gep_0\novx|_{x=x_0}=\xi\nov
\ee
for such a point $x_0$ that $g(x_0)=1$. Therefore $hu(1;1|4)$
is indeed the global symmetry algebra that leaves invariant the vacuum
solution. It contains the $N=2$ global conformal supersymmetry algebra
spanned by the generators (\ref{confg}), (\ref{oddg}), (\ref{eg}).

As usual, the gravitational fields (i.e., frame and Lorentz connection)
are associated with the generators of translations and Lorentz rotations
in the Poincare or AdS subalgebras of the conformal algebra.
For the
flat Minkowski space one can choose
\be
\go_0 =\go_f=\frac{1}{2}d x^\un \gs_\un{}^{\ga\gb}a_\ga a_\gb
\label{fc}
\ee
thus setting the Lorentz connection equal to zero.
Here $\gs_\un{}^{\ga\gb} = \gs_n{}^{\ga\gb}$.
The function $g_f$ that gives rise to the flat gravitational
field (\ref{fc}) is
\be
g_f=\exp(-\frac{x^{\ga\gb}}{2}a_\ga a_\gb)\,,
\label{gfc}
\ee
where we use the notation
\be
x^{\ga\gb}=x^\un\gs_\un{}^{\ga\gb}\,,
\qquad x^\un=\frac{1}{2}\gs^\un{}_{\ga\gb}x^{\ga\gb}\,.
\ee

Within the oscillator realization of the $3d$ conformal algebra
(\ref{confg}), the embedding of the
 $AdS_3$ algebra $o(2,2)\subset o(3,2)$ can be realized as follows
\be
\L_\ga{}^\gb=a_\ga a^{+\gb}-\frac{1}{2}\gd_\ga^\gb a_\gga a^{+\gga}\,,
\label{adstr}
\ee
$$ \P_{\ga\gb}=\frac{1}{2}(a_\ga a_\gb +
\frac{\gl^2}{4}a^+_\ga a^+_\gb)\,. $$
The $AdS_3$ gravitational fields are identified with the
1-forms taking values in the $AdS_3$ algebra
\be
\go_0 =\go_{AdS_3}=dx^\un\Big(\frac{1}{2} e_\un{}^{\ga\gb} (x)
(a_\ga a_\gb+\frac{\gl^2}{4}a^+_\ga
a^+_\gb)+\go_\un{}^\ga{}_\gb (x)
(a_\ga a^{+\gb}-\frac{1}{2}\gd_\ga^\gb a_\gga a^{+\gga})\Big) \,,
\label{adscon}
\ee
where $e_\un{}^{\ga\gb} (x)$ and $\go_\un{}^\ga{}_\gb (x)$
are, respectively, the  dreibein and Lorentz connection of $AdS_3$.
A particular choice of the $AdS_3$ gravitational fields that
solves the vacuum equations (\ref{ezc}) and corresponds to the
``stereographic" coordinates of $AdS_3$ is

\be
e_\un{}^{\ga\gb}=\frac{4}{(4+\gl^2x^2)^2}
\Big((4-\gl^2 x^2)\gd_\un^\uk+4\gl x^\um
\gep_{\un\um}{}^\uk+2\gl^2
x_\un x^\uk\Big)\gs_\uk{}^{\ga\gb}\,,
\label{eads}
\ee
and
\be
\go_\un{}^\ga{}_\gb=-\frac{\gl}{2}e_\un{}^{\ga}{}_{\gb}\,,
\label{wads}
\ee
where
\be
x^2 =\eta^{\un\um}x_\un x_\um=\frac{1}{2}x^{\ga\gb}x_{\ga\gb}\,.
\ee
The metric tensor of the $AdS_3$ is
\be
g_{\un\um}=\frac{1}{2}e_\un{}^{\ga\gb}e_{\um\ga\gb}=
16\frac{\eta_{\un\um}}{(4+\gl^2x^2)^2}\,.
\ee
Let us mention that although the metric tensor is
built {}from $\go_0$ and, therefore, is
invariant under the global symmetry transformations
these symmetries are not necessarily associated with
the Killing vectors of the metric tensor.
This is only true for the Lorentz rotations
and (Poincare or AdS) translations.

It is elementary to see that
the representation (\ref{wvg}) for the $AdS_3$
vacuum fields (\ref{adscon}) - (\ref{wads}) is provided with
the gauge function
\be
g_{AdS_3}=\frac{1}{2}\sqrt{4+\gl^2x^2}
\exp \left( -\frac{x^{\ga\gb}}{2}
(a_\ga a_\gb+\frac{\gl^2}{4}a^+_\ga a^+_\gb-
\gl a^+_\ga a_\gb)\right) \,,
\label{gf}
\ee
having the inverse
\be
g_{AdS_3}^{-1}=\frac{1}{2}\sqrt{4+\gl^2x^2}
\exp \left( \frac{x^{\ga\gb}}{2}
(a_\ga a_\gb+\frac{\gl^2}{4}a^+_\ga a^+_\gb-
\gl a^+_\ga a_\gb)\right) \,.
\label{gmpf}
\ee
Note that in the flat limit $\lambda \to 0$ one recovers the flat
gauge function (\ref{gfc}).

\section{$3d$ Conformal Field Equations}
It was shown that the equations of motion for
massless \cite{unf} and massive \cite{BPV} fields in ${\rm AdS}_3$ admit a
formulation in terms of generating functions
\be
\label{fe}
C(y|x)=\sum_{l=0}\frac{1}{l!}C(x)_{\ga_1\ldots\ga_l}y^{\ga_1}\cdots
y^{\ga_l}
\ee
with the auxiliary spinor variables $y^\ga$. The flat limit of the
free equations of motion for the scalar and spinor massless fields
of \cite{unf} has the form
\be
\label{deq}
dC(y|x)=\frac{1}{2}dx^\un
\gs_\un{}^{\ga\gb}\frac{1}{\ptl y^\ga\ptl y^\gb}C(y|x)
\,.
\ee
This equation decomposes into two independent subsystems
for even functions \hfill\\ $C_e (-y|x)=C_e (y|x)$ and odd functions
$C_o (-y|x)=-C_o (y|x)$ which describe the massless
scalar and spinor respectively.

In \cite{V} the spinor variables $y^\ga$ were interpreted
as generating elements of the ${\rm AdS}_3$ higher spin algebra while the
0-form $C(y|x)$ took its values in the so-called twisted adjoint
representation of this algebra. In  \cite{PV} it was then shown that
an appropriate modification of the formulation of \cite{V} allows for
a uniform description of both  massless and massive matter fields.
This formulation does not make manifest the conformal symmetries
expected for the massless case, however. The key observation of this paper
is
that for the massless case the same equations (\ref{deq})  admit a
different realization in the Fock space (\ref{fr}) that makes the
higher spin conformal symmetries of the system manifest.

Let us introduce the Fock-space vector
\be
|\Phi(a^+|x) \rangle = C(a^+ |x )* \vac \,,\qquad
C(a^+|x) = \sum_{l=0}^\infty \frac{1}{l!}c_{\ga(l)}(x)
\underbrace{a^{+\ga}\cdots a^{+\ga}}_l\,,
\label{cdef}
\ee
where 0-forms $c_{\ga(l)}(x)$ are totally symmetric
multispinors\footnote{We follow the conventions
of \cite{V2} convenient for the component analysis
of complicated tensor structures: upper and lower indices denoted by the
same letter should be first separately symmetrized and then the maximal
possible number of them should be contracted;
a number of indices can be indicated in brackets by writing e.g.
$\ga(l)$ instead of repeating $l$ times the index $\ga$.}.
The system of equations
\be
d|\Phi \rangle -\go  * |\Phi \rangle =0
\label{ffe}
\ee
concisely encodes Klein-Gordon and Dirac equations for
the scalar field $c(x)$ and spinor field $c_\ga(x)$
provided that
the equation (\ref{ezc}) that guarantees the formal consistency
of (\ref{ffe}) is true. Indeed, the
choice $\go=\go_f$ in the form (\ref{fc}) makes the equation
(\ref{ffe}) equivalent to (\ref{deq}).  Let us note
that the equations on the component fields
$c_{\ga(n)} (x)$ are Lorentz and scale invariant
as a consequence of the Lorentz invariance (\ref{linv})
and definite scaling (\ref{conv}) of the vacuum $\vac$.

Recall that, as shown in \cite{unf} for the ${\rm AdS}_3$ case,
the meaning of the equations (\ref{deq}) is that they impose
the dynamical equations on the massless matter fields identified with
the rank-0 and rank-1 spinors
\be
c(x) = C(0 |x)\,,
\label{cx}
\ee
\be
 c_\ga (x) = \frac{\partial}{\partial a^{+\ga} }
C(a^+ |x) \Big |_{a^{+\ga} = 0}\,
\label{cax}
\ee
 and express all highest spinors {}from (\ref{cdef}) via higher order
derivatives of $c(x)$ and $c_\ga (x)$.
In the simplest case of the flat space this can be seen as follows.
Substituting $\go_f$ into (\ref{ffe}) we obtain
\bee
&\ptl_\un c_{\ga(l)}-\frac{1}{2}\gs_\un{}^{\ga\ga}c_{\ga(l+2)}=0\,.
\label{ffs}
\eee
The first two equations for even $l$, i.e. for $l=0$ and $l=2$, are
\be
\ptl_\un c-\frac{1}{2}\gs_\un{}^{\ga\ga}c_{\ga(2)}=0\,,
\label{ek0}
\ee
\be
\ptl_\un c_{\ga(2)}-\frac{1}{2}\gs_\un{}^{\ga\ga}c_{\ga(4)}=0\,.
\label{ek2}
\ee
{}From (\ref{ek0}) we derive that $c_{\ga(2)}=\gs^\un{}_{\ga\ga}\ptl_\un c$.
Substituting it into (\ref{ek2}) and
multiplying by $\gs^\um{}_{\gb\gb}$ we obtain
\be
\gs^\un{}_{\ga\ga}\gs^\um{}_{\gb\gb}\ptl_\un\ptl_\um c=c_{\ga(2)\gb(2)}\,.
\ee
The condition that $c_{\ga(2)\gb(2)}$ is totally symmetric is equivalent
to the Klein-Gordon equation for~$c(x)$
\be
\Box c(x)=0\,.
\label{kg}
\ee
The equations for all other even values of
$l$ impose no further differential
equations on $c(x)$
just expressing  highest multispinors via the highest derivatives
of the $c(x)$. It is straightforward to see that the resulting expression
for the even part $C_e$ of $C$ acquires the form
\be
C_e (a^+|x)=\sum_{q=0}^\infty\frac{1}{(2q)!}
\gs^{\un_1}{}_{\ga\ga}\cdots\gs^{\un_q}{}_{\ga\ga}
\ptl_{\un_1}\cdots\ptl_{\un_q}c(x)
\underbrace{a^{+\ga}\cdots a^{+\ga}}_{2q}\,.
\label{ce}
\ee

The situation with the fermion is analogous.
Starting {}from the equation (\ref{ffs}) for $l=1$ we obtain
\be
c_{\ga\gb(2)}=\gs^\un{}_{\gb\gb}\ptl_\un c_\ga\,.
\ee
Again, the condition that the third rank multispinor
$c_{\ga\gb(2)}$ is symmetric implies the
Dirac equation for $c_\ga(x)$
\be
\gs^{\un\gb}{}_\ga\ptl_\un c_\gb (x)=0\,.
\label{dirac}
\ee
All other equations for odd values of
$l$ express highest spinors via derivatives
of $c_\ga (x)$.
For the odd part $C_o$ of $C$ we get
\be
C_o(a^+|x)=\sum_{q=0}^\infty\frac{1}{(2q+1)!}
\gs^{\un_1}{}_{\ga\ga}\cdots\gs^{\un_q}{}_{\ga\ga}\ptl_{\un_1}
\cdots\ptl_{\un_q}c_\ga(x)
\underbrace{a^{+\ga}\cdots a^{+\ga}}_{2q+1}\,.
\label{co}
\ee
We therefore conclude that, for the flat
connection $\go_f$, the system (\ref{ffe}) is equivalent to
\be
\!\!|\Phi(a^+|x)\rangle\!=\!\sum_{q=0}^\infty\frac{1}{(2q)!}
\gs^{\un_1}{}_{\ga\ga}\cdots\gs^{\un_q}{}_{\ga\ga}
\ptl_{\un_1}\cdots\ptl_{\un_q}
\Big( c (x)+\frac{1}{2q+1}c_\ga (x) a^{+\ga} \Big)
\underbrace{a^{+\ga}\cdots a^{+\ga}}_{2q}*\vac
\label{c}
\ee
along with the dynamical equations (\ref{kg}) and (\ref{dirac}).

Analogously, for $\go_{AdS_3}$
the system (\ref{ffe}) is equivalent to the
massless Klein-Gordon
and Dirac equations in $AdS_3$
\be
(g^{\un\um}D_\un D_\um-\frac{3}{4}\gl^2)c(x)=0\,,
\label{adscg}
\ee
\be
e^{\un\gb}{}_\ga D_\un c_\gb(x) =0
\label{adsd}
\ee
along with
\be
\!\!|\Phi(a^+|x)\rangle\!=\!\sum_{q=0}^\infty\frac{1}{(2q)!}
e^{\un_1}{}_{\ga\ga}\cdots e^{\un_q}{}_{\ga\ga} D_{\un_1}\cdots D_{\un_q}
\Big (c(x)+\frac{1}{2q+1}c_\ga(x) a^{+\ga}\Big)
\underbrace{a^{+\ga} \cdots  a^{+\ga}}_{2q}*\vac\,,
\label{adsc}
\ee
where $D_\un$ is the $AdS_3$ Lorentz covariant derivative
\be
D_\un c_{\ga(l)}=\ptl_\un c_{\ga(l)}+
l\go_\un{}^\gb{}_\ga c_{\ga(l-1)\gb}\,,
\ee
\be
D_\un e_\um^{\ga\ga}-\un\leftrightarrow \um=0\,.
\ee
Note that, as usual, we identify  massless scalar
and spinor in $AdS_3$ with the conformal fields.
With this convention, the massless scalar field in $AdS_d$
has a non-zero mass-like term
$m_0^2=-(\gl^2 /4)d(d-2)$ being in agreement with
(\ref{adscg}) for $d=3$.

\section{3d Conformal Higher Spin Symmetries}
\label{Conformal Higher Spin Symmetries if $3d$ Massless Equations}

The system of equations (\ref{ezc}) and (\ref{ffe}) is
invariant under the infinite-dimensional local
conformal higher spin symmetries of the form
(\ref{gtw}) and
\be
\gd |\Phi\rangle =\gep* |\Phi\rangle
\label{gtc}
\ee
($\gep = \gep (a^+ ,a,k |x)$).
Once a  particular vacuum solution
$\go=\go_0$ is fixed,
the local higher spin symmetry (\ref{gtc}) breaks down to the global
higher spin symmetry (\ref{e0}).
Therefore the system (\ref{ffs})
and its  $AdS_3$ analog are invariant under the
infinite-dimensional algebra $hu(1;1|4)$
of global 3d conformal higher spin symmetries
\be
\label{gs}
\gd |\Phi\rangle=\gep_0*|\Phi\rangle\,,
\ee
where $\gep_0$ satisfies the equation (\ref{ipgs}) with the flat
or $AdS_3$ connection (\ref{fc}) or (\ref{adscon}).
Once the higher components $c_{\ga (l)}$ are expressed via
higher derivatives of the dynamical spin zero and spin 1/2 fields
by (\ref{c}) and (\ref{adsc}), this implies the invariance of the
massless Klein-Gordon and Dirac equations (\ref{kg}), (\ref{dirac}) and
(\ref{adscg}), (\ref{adsd})
under the conformal higher spin symmetries.
Thus, the fact that massless Klein-Gordon and
Dirac equations are reformulated
in the form of the flatness conditions  (\ref{ffe}) and
zero-curvature equation (\ref{ezc}) makes
higher spin conformal symmetries of these equations manifest.

The $N=2$ conformal SUSY algebra $osp(2|4,\R)$ spanned by the elements
(\ref{confg}), (\ref{oddg}) and (\ref{eg}) is a
finite-dimensional subalgebra of the $3d$
conformal higher spin symmetry algebra $hu(1;1|4)$.
Including the $u(1)$ factor generated by
the constant element in $hu(1;1|4)=u(1)\oplus hsu(1;1|4)$,
$u(1)\oplus osp(2|4,\R)$ forms a maximal finite-dimensional
subalgebra of $hu(1;1|4)$.
Note that because of (\ref{c}) or (\ref{adsc}) and of
the quantum-mechanical nonlocality of the star product
(\ref{ints}), the higher degree of $\gep_0 (a^+ ,a,k|x)$ as a
polynomial of $a_\ga$ and $a^{+\gb}$ is the higher
space-time derivatives appear in the higher spin conformal
transformations. This is a particular manifestation of the well
known fact that the higher spin symmetries mix higher derivatives of
the dynamical fields.

The explicit form of the transformations can be obtained by
the substitution of (\ref{c})
((\ref{adsc}) in $AdS_3$ case) into (\ref{gs}). In practice, it is
most convenient to evaluate the higher spin conformal transformations
for the generating parameter
\be
\label{xi}
\xi^j (a^+,a,k;h^+ , h )=\xi
\exp(a^{+\ga}h_\ga+a_\ga h^{+\ga})(k)^j\,,
\ee
where $\xi$ is an infinitesimal parameter and
$h_\ga$, $h^{+\ga}$ are spinor ``sources". The polynomial symmetry
parameters can be obtained via differentiation of
$\xi^j (a^+,a,k;h^+ , h )$
with respect to
$h_\ga$ and $h^{+\ga}$.
Using (\ref{e0}), (\ref{gf}) and the star product
(\ref{ints}) we obtain upon evaluation of
elementary Gaussian integrals
\bee
&\gep_0(a^+,a,k;h^+ , h|x) =\xi
\exp\left ( \frac{1}{8+2\gl^2x^2}
\Big (x^{\ga\ga}(8a_\ga h_\ga+4\gl a_\ga h^+_\ga-
4\gl a^+_\ga h_\ga-2\gl^2 a^+_\ga h^+_\ga)+ \right. \nn\\
&\left.
{}+8a^{+\ga}h_\ga+8a_\ga h^{+\ga}+4\gl x^2a^\ga h_\ga+
\gl^3 x^2 a^+_\ga h^{+\ga}\Big) \right ) (k)^j\,.
\label{ez}
\eee

Substitution of $\gep_0$ into (\ref{gs}) gives
 the global higher spin conformal symmetry transformations
induced by the parameter $\xi^j(a^+,a,k;h^+,h)$ (\ref{xi})
\bee
&\delta |\Phi(a^+|x)\rangle=\xi
\exp \Big ( \frac{1}{8+2\gl^2 x^2}\Big (
x^{\ga\ga}(4h_\ga h_\ga-\gl^2 h^+_\ga h^+_\ga -4\gl h_\ga a^+_\ga-
2\gl^2 h^+_\ga a^+_\ga)+\nn\\
&{}+8a^{+\ga}h_\ga+\gl^3 x^2 a^+_\ga h^{+\ga}+
(4-\gl^2 x^2)h^{+\ga}h_\ga\Big)\Big)\nn\\
&C\Big((-1)^j\Big(a^{+\ga}+\frac{1}{8+2\gl^2 x^2}(
8x^{\ga\gb} h_\gb+4\gl x^{\ga\gb} h^+_\gb+8h^{+\ga}
-4\gl x^2 h^\ga )\Big)\Big| x\Big)*\vac \,.
\label{dc}
\eee
Setting $\gl=0$ we obtain the  flat space   formula
\be
\!\!\delta | \Phi(a^+|x)\rangle\!=\!\xi
\exp\! \Big(\frac{1}{2}x^{\ga\ga}h_\ga h_\ga+
a^{+\ga}h_\ga +\frac{1}{2}h^{+\ga}h_\ga \Big)
C\Big((-1)^j(a^{+\ga}+x^{\ga\gb}h_\gb+h^{+\ga})\Big| x\Big) *\vac\,.
\label{fdc}
\ee

Differentiating with respect to the sources $h_\ga $ and $h^{+\ga}$
one derives explicit expressions for the particular
global higher spin conformal transformations.
For the transformations of the dynamical fields
(\ref{cx}), (\ref{cax}) these expressions further simplify.
For example, the bosonic transformation of the dynamical scalar
in the flat space is
\be
\delta c=\xi
\exp \Big(\frac{1}{2}x^{\ga\ga}h_\ga h_\ga+
\frac{1}{2}h^{+\ga}h_\ga \Big)
C\Big(x^{\ga\gb}h_\gb+h^{+\ga}\Big| x\Big)
\ee
with (\ref{ce}) substituted into the right hand side.
Let us stress that such a compact form of the higher spin
conformal transformations is a result of the
reformulation of the dynamical equations in the unfolded form of the
covariant constancy conditions, i.e.
in terms of a flat section of the Fock bundle.
Note that a ``brutal force'' search of the higher spin conformal
transformations quickly gets very complicated (especially in the
$AdS_3$ case).

For at most quadratic
conformal superalgebra generators (\ref{confg}),
(\ref{oddg}), (\ref{eg}) one immediately obtains
in the flat space
$$
\L_\ga{}^\gb\fv=\Big( x^{\gb\gga}
\frac{\ptl^2}{\ptl a^{+\ga}\ptl a^{+\gga}}+
a^{+\gb}\frac{\ptl}{\ptl a^{+ \ga}}-
\frac{1}{2}\gd_\ga^\gb\Big ( x^{\gga\gga}
\frac{\ptl^2}{\ptl a^{+\gga}\ptl a^{+\gga}}+
a^{+\gga}\frac{\ptl}{\ptl a^{+\gga}}\Big ) \Big)\fv\,,
$$
$$
\P_{\ga\ga}\fv=\frac{1}{2}
\frac{\ptl^2}{\ptl a^{+\ga}\ptl a^{+\ga}}\fv\,,
$$
$$
\K^{\ga\ga} \fv=\frac{1}{2}
\Big (x^{\ga\ga}+a^{+\ga} a^{+\ga}+2x^{\ga\gb} a^{+\ga}
\frac{\ptl}{\ptl a^{+\gb}}+x^{\ga\gb} x^{\ga\gb}
\frac{\ptl^2}{\ptl a^{+\gb}\ptl a^{+\gb}}\Big ) \fv\,,
$$
\be
\label{scsa}
\D \fv=\frac{1}{2}\Big(1+a^{+\ga}\frac{\ptl}{\ptl a^{+\ga}}+
x^{\ga\ga}\frac{\ptl^2}{\ptl a^{+\ga}\ptl a^{+\ga}}\Big)\fv\,,
\ee
$$
\Q_{j\ga}\fv=\frac{(i)^j}{\sqrt 2}
\frac{\ptl}{\ptl a^{+\ga}}|\Phi((-1)^ja^+|x)\rangle\,,
$$
$$
\S_j{}^\ga\fv=\frac{(i)^j}{\sqrt 2}
\Big(a^{+\ga}+x^{\ga\gb}
\frac{\ptl}{\ptl a^{+\gb}}\Big)|\Phi((-1)^ja^+|x)\rangle\,,
$$
$$
\J \fv=\frac{1}{4i}|\Phi(-a^+|x)\rangle\,.
$$
Taking into account that the dynamical equations for
$|\Phi(a^+|x)\rangle$
have the form (\ref{deq}) one can replace all second derivatives in
$a^{+\ga}$ by the space-time derivatives. This leads to the following
standard expressions
$$
\L_{nm}\fv=\Big(x_m\ptl_n-x_n\ptl_m+
\frac{1}{2}\gep_{nmk}\gs^{k\ga}{}_\gb a^{+\gb}\frac{\ptl}
{\ptl a^{+\ga}}\Big)\fv\,,
$$
\be
\P_n\fv=\ptl_n \fv\,,
\ee
$$
\K_{n} \fv= \Big(x_n+2x_nx^k\ptl_k-x^2\ptl_n
+\frac{1}{2}\gs_{n\ga\ga}(a^{+\ga} a^{+\ga}+2x^{\ga\gb} a^{+\ga}
\frac{\ptl}{\ptl a^{+\gb}})\Big)\fv\,,
$$
$$
\D \fv=\Big(x^k \ptl_k+\frac{1}{2}+\frac{1}{2}a^{+\ga}\frac{\ptl}
{\ptl a^{+\ga}}\Big)\fv\,.
$$

For the dynamical scalar and spinor fields
(\ref{cx}) and (\ref{cax}) we find the expected results
\bee
&&\L_{nm}c=(x_m\ptl_n-x_n\ptl_m)c\,,\nn\\
&&\P_n c=\ptl_n c\,,\nn\\
&&\K_n c=(x_n+2x_n x^k\ptl_k-x^2\ptl_n )c\,,\nn\\
&&\D c=(\frac{1}{2}+x^k \ptl_k) c\,,\\
&&\Q_{j\ga}c=\frac{(-i)^j}{\sqrt 2}c_\ga\,,\nn\\
&&\S_j^\ga c=\frac{(-i)^j}{\sqrt 2}
x^{\ga\gb} c_\gb\,,\nn\\
&&\J c=-\frac{i}{4}c\,,\nn
\eee
\bee
&&\L_{nm}c_\ga=\Big((x_m\ptl_n-x_n\ptl_m)\gd_\ga^\gb+
\frac{1}{2}\gep_{nmk}\gs^{k\gb}{}_\ga\Big) c_\gb\,,\nn\\
&&\P_n c_\ga=\ptl_n c_\ga\,,\nn\\
&&\K_n c_\ga= \Big((2x_n+2x_n x^k \ptl_k-x^2\ptl_n)\gd_\ga^\gb-
\gep_{nmk}x^m\gs^{k\gb}{}_\ga\Big) c_\gb\,,\nn\\
&&\D c_\ga=(1+x^k\ptl_k) c_\ga\,,\\
&&\Q_{j\ga}c_\gb=\frac{(i)^j}{\sqrt 2}
\gs^n{}_{\ga\gb}\ptl_n c\,,\nn\\
&&\S_j^\ga c_\gb=\frac{(i)^j}{\sqrt 2}
\Big((1+x^k\ptl_k)\delta_\gb^\ga+
x_n\ptl_m\gep^{nmk}\gs_k{}^\ga{}_\gb\Big)c\,,\nn\\
&&\J c_\ga=\frac{i}{4}c_\ga\,.\nn
\eee

Note that the particular form of the dependence on the
space-time coordinates $x^{\ga\gb}$ originates {}from the choice of
the gauge function (\ref{gfc}). The approach we use is applicable
to any other coordinate system
and conformally flat background (for example, ${\rm AdS}_3$).

\section{Field Theory - Singleton Duality Via Bogolyubov
Transform}

The formulation of the relativistic  higher spin dynamics
proposed in this paper operates
in terms of the Fock module $F$ (\ref{akv}) defined
with respect to auxiliary oscillators associated with
the 3d conformal superalgebra $osp(2|4,\R)$ via
(\ref{confg}), (\ref{oddg}) and (\ref{eg}).
This Fock module is analogous to the Fock
representations of $sp(4,\R)$ identified with the Dirac singletons
\cite{Dirac}.
The difference is that the singleton representation $S$ is unitary
while the Fock module $F$ is not.
In this section we show that our approach makes
the duality between the two
types of representations manifest by virtue of a certain Bogolyubov
transform.
This parallelism extends far enough. In particular,
the Flato-Fronsdal theorem \cite{FF} that the tensor product
of the two singleton representations is equivalent to the direct sum
of all $AdS_4$ massless representations acquires a simple dynamical
interpretation in the unfolded formulation of the higher spin dynamics.

The fact that the involution $\mu$ (\ref{inv}) maps the
oscillators $a_\ga$ and $a^{+\ga}$ to themselves
implies that the Fock module
(\ref{cdef}) is not unitary.
This is in agreement with the
fact that the vacuum $\vac$ is Lorentz invariant (\ref{linv})
and, as a result, the module (\ref{cdef}) decomposes into the
infinite sum of the finite-dimensional representations of the
noncompact
$3d$ Lorentz algebra $o(2,1)$ identified with the component fields
$c_{\ga (l)} (x)$. (Recall that noncompact algebras do not admit
finite-dimensional unitary representations.)

The unitary Fock module of $sp(4,\R)$ is built in terms of the
 oscillators
\be
\label{uosc}
[\hat b_i^\pm \,,\hat b_j^\pm ] =0\,,
\qquad  [\hat b_i^- \,,\hat b_j^+ ] = \delta_{ij}\,,
\qquad i,j=1,2\,,
\ee
satisfying the Hermitian conjugation conditions
\be
\label{bog1}
(\hat b_i^\pm )^\dagger = \hat b_i^\mp \,.
\ee
The corresponding Fock vacuum $\uvac$ is defined according to
\be
\hat b^-_i \uvac = 0\,, \qquad \uvac \hat b^+_i = 0\,.
\ee
The unitary left and right Fock modules built {}from the vacuum $\uvac$
identify with the supersingleton $S$ and its conjugate $\bar{S}$,
respectively. (The supersingleton $S$
decomposes into two irreducible representations
of the $sp(4,\R)$ associated with the subspaces built
{}from $\uvac$ with the
aid of even and odd numbers of creation operators, called
Rac and Di, respectively \cite{FF}.)

The relationship between the two sets of oscillators is
\bee
\label{bog2}
\hat b_j^\pm = \frac{1}{\sqrt{2}} \Big (
\hat a_j \pm \hat  a^{+j} \Big )\,,
\eee
\be
\label{bog3}
\hat a_j =\frac{1}{\sqrt{2}} \Big ( \hat b^+_j +\hat b^-_j )\,,\qquad
\hat a^{+j} =\frac{1}{\sqrt{2}} \Big ( \hat b^+_j -\hat b^-_j )\,.
\ee
The unitary Fock vacuum can be realized  in terms of the
star product algebra (\ref{ints}) as
\be
\uvac =2(1+k)\exp\Big (-\gd^{ij}(a_i a_j-a^+_i a^+_j)\Big )\,.
\ee
 We therefore conclude that, in our approach,
there is a natural duality between the field-theoretical module
$F$ used in the unfolded formulation of the 3d conformal dynamics
and unitary module $S$.

The idea that some duality of this kind takes place has been
worked out earlier \cite{Gun} for the $4d$ case
and is very interesting in the context of the AdS/CFT
correspondence. The new point is that
it takes a simple form of the Bogolyubov transform
(\ref{bog2}), (\ref{bog3}) in the framework of the unfolded
formulation of
the 3d conformal dynamics. Let us stress that the dependence
on the space-time coordinates of the elements of the field
$|\Phi (x) \rangle $ is determined completely by the equation
(\ref{ffe}) in terms of its value at any fixed point $x_0$. This means
that the module $|\Phi (x_0) \rangle $ contains the complete
information on the on-mass-shell dynamics of the 3d conformal fields
analogously to the fact that the singleton module contains the complete
information on the (on-mass-shell) quantum states of the free theory.
We believe that this phenomenon is quite general and the
unfolded formulation of the dynamical systems in the form of some
flatness (i.e., covariant constancy or zero-curvature) conditions
will make the duality between the classical and quantum description
of the dynamical systems manifest for general case.

The duality between unitary and unfolded formulations  of the
dynamical systems exhibiting higher spin symmetries admits an
interesting extension to
the $AdS_4$ higher spin gauge theories.
As shown by Flato and Fronsdal \cite{FF} the tensor product of
singleton representations amounts to the direct sum of the
unitary representations of $sp(4,\R)$ associated with the massless
fields of all spins in $AdS_4$. In particular,
a $AdS_4$ massless field of every spin
$s= 0,1/2,1,3/2, \ldots$ appears in two copies
in $S\otimes \bar{S}$. (See also \cite{KV0}
for the straightforward analysis). Remarkably, this is exactly
the spectrum of the simplest
(i.e., without non-Abelian Yang-Mills gauge symmetries)
supersymmetric $AdS_4$ higher spin gauge theory built in
\cite{Ann,more} based on $hu(1;1|4)$.
In \cite{KV0} it was shown that this fact implies that
$hu(1;1|4)$ admits a unitary representation to be associated with
the one-particle states of the quantum higher spin gauge theory.
In \cite{KV1} it was then shown that all consistent $4d$ higher
spin gauge theories with various Yang-Mills groups admit unitary
representations associated with certain tensor products of pairs of
singletons.

Now, let us follow the field-theoretical picture.
A particular basis in the tensor product $F^l \otimes {F}^r$
of the left and right
Fock modules is spanned by the elements of the form
\be
\label{bel}
a^+_{\ga_1}* \ldots *a^+_{\ga_n}* \vac *
a_{\gb_1}* \ldots * a_{\gb_m} \,.
\ee
$F^l \otimes {F}^r$ can be identified with the algebra
of endomorphisms of $F$. The
star product algebra (\ref{ints}) also can be interpreted as
the algebra of linear operators in $F$
and therefore it can be identified with
$F^l \otimes {F}^r$.\footnote{Here
we discard the normalizability issues.
Note that an operator (\ref{bel})
in $F$ is described by the infinite matrix having a finite number
of non-zero elements, while the polynomial elements of the star product
algebra have the Jacobi form with an infinite number of non-zero elements
but at most a finite number of non-zero diagonals. This means that
a polynomial in the star product algebra is described by an infinite sum
in the basis (\ref{bel}).  The questions
of how well defined are the corresponding infinite sums
sometimes become important.}
The point is that the $AdS_4$ higher spin gauge fields
are the gauge fields (1-forms) taking values in this algebra.
This provides
an interesting dynamical realization of the Flato-Fronsdal theorem.
Moreover, it has been shown recently \cite{kvz}
that, in agreement with the AdC/CFT correspondence
\cite{AdS/CFT}, the set of the
$AdS_4$ higher spin gauge fields is in one-to-one correspondence
with the set of 3d conformal higher spin currents.

The important question of the explicit form of the
AdS/CFT correspondence
in the framework of the higher spin gauge theories
requires a more detailed analysis of the higher spin dynamics
and will be considered elsewhere. One relevant issue is that
the Fock-module realization of the 3d equations
proposed in this paper is different {}from the approach to
the nonlinear $3d$ higher spin theory given in \cite{V,PV} where the
equation (\ref{ffe}) was formulated in terms of the twisted
adjoint representation of the $AdS_3$ higher spin algebra realized in
terms of the smaller set of oscillators $y^\ga$ according to (\ref{fe}).
As the formulation of \cite{V,PV} works both for massless \cite{V} and for
massive matter fields \cite{PV} it has no manifest evidence of the
conformal higher spin symmetries. The approach
developed in this paper therefore raises an
important problem of the search
of a manifestly conformal theory
describing higher spin interactions
of 3d massless matter, based on the Fock modules rather than on the twisted
adjoint representation. Let us note that this alternative is
expected to be analogous to the
realization of the 2d higher spin matter system of \cite{d2}
originally formulated in terms of the Fock module. Also it is tempting
to speculate that the resulting models may have some relationship with
the models of non-commutative solitons
\cite{Gopuk,sol} discussed recently in the
context of the noncommutative phase of  superstring theory \cite{nc},
which are realized in terms of Fock modules with respect to the
non-commutative space-time coordinates.

\section*{Acknowledgements}
M.V. would like to thank I.~Bars, M.~Gunaydin and E.~Witten for useful
discussions.
This research was supported in part by INTAS, Grant No.99-1-590,
the RFBR Grant No.99-02-16207 and the RFBR Grant No.00-15-96566.
M.V. would like to thank the
CIT-USC Center for Theoretical Physics for hospitality where some part
of this work was performed.

\end{document}